\documentstyle[preprint,tighten,pra,aps]{revtex}
\begin{document} \draft

\title{\LARGE \bf WEINBERG'S  APPROACH AND ANTISYMMETRIC TENSOR
FIELDS\thanks{Presented at the VII Wigner Symposium, College Park,
MD, USA, August 24-29, 2001.}}

\author{Valeri V. Dvoeglazov}

\address{Universidad de Zacatecas, Apartado Postal 636\\
Suc. UAZ, Zacatecas 98062 Zac., M\'exico\\
E-mail: valeri@ahobon.reduaz.mx\\
URL: http://ahobon.reduaz.mx/\~~valeri/valeri.htm}

\date{August 31, 2001}

\maketitle

\medskip

\begin{abstract}
We extend the previous series of articles~\cite{HPA}
devoted to finding mappings between the Weinberg-Tucker-Hammer
formalism and antisymmetric tensor fields. Now we take into account
solutions of different parities of the Weinberg-like equations.
Thus, the Proca, Duffin-Kemmer and Bargmann-Wigner formalisms
are generalized.
\end{abstract}

\vspace*{1cm}

The content of my talk is the following:

\begin{itemize}

\item
Introductory notes on the field theory in the $2(2J+1)$-component
formalism presented in the famous papers~\cite{Wein}  ``Feynman Rules for
Any Spin".

\item
The group approach. Wigner and Weinberg. The little group, group
contraction  and  Maxwell-like formalism.

\item
The mapping between the Weinberg formulation  and the second-rank
antisymmetric tensor formulation.

\item
Various types of solutions with different parity properties. Particular
cases.

\item
Massless limits.

\item
The Bargmann-Wigner formalism for symmetric spinors of the $2J$ rank.

\item
Derivation of the Proca theory (the first case in the
Weinberg-Tucker-Hammer (WTH) formalism for $J=1$).

\item
Modifications of the Dirac equation (the parity-violating
version and the Barut second-order equation).

\item
Modifications of the Bargmann-Wigner formalism.

\end{itemize}

I began to study these issues in the works for my M. Sc. thesis, published
in~\cite{old}.  In ref.~\cite{Dv-rev} I reviewed the Weinberg
$2(2J+1)$-component formalism~\cite{Wein}.  It is based on the following
postulates:

\begin{itemize}

\item
The fields transform according to the formula:
\begin{equation}
U [\Lambda, a] \Psi_n (x) U^{-1} [\Lambda, a] = \sum_m D_{nm}
[\Lambda^{-1}] \Psi_m (\Lambda x +a)\,,\label{1} \end{equation}
where $D_{nm} [\Lambda]$ is some representation of $\Lambda$; $x^\mu \rightarrow
\Lambda^\mu_{\quad\nu} \,\,x^\nu +a^\mu$, and $U [\Lambda, a]$ is a
unitary operator.

\item
For $(x-y)$ spacelike one has
\begin{equation}
[\Psi_n (x), \Psi_m (y) ]_\pm =0\,\label{2}
\end{equation}
for fermion and boson fields, respectively.

\item
The interaction Hamiltonian density is said by S. Weinberg to be a scalar,
and it is constructed out of the creation and annihilation operators for
the free particles described by the free Hamiltonian $H_0$.

\item
The $S$-matrix is constructed as an integral of the $T$-ordering
product of the interaction Hamiltonians by the Dyson's formula.

\end{itemize}

In this talk we shall be mainly interested in the free-field theory.
Weinberg wrote: ``In order to discuss theories with parity conservation it
is convenient to use $2(2J+1)$-component fields, like the Dirac field.
These do obey field equations, which can be derived as\ldots consequences
of (\ref{1},\ref{2})."\,\footnote{In the $(2J+1)$ formalism fields
obey only the Klein-Gordon equation, according to the Weinberg wisdom.
See, however, ref.~\cite{Br-Feyn}.} In such a way he proceeds to form the
$2(2J+1)$-component object
$$\Psi =\pmatrix{\Phi_\sigma\cr \Xi_\sigma\cr}$$
transforming according to the Wigner rules. They are the following ones:
\begin{mathletters}
\begin{eqnarray}
\Phi_\sigma ({\bf p}) &=& \exp (+\Theta \,\hat {\bf p} \cdot {\bf J})
\Phi_\sigma ({\bf 0}) \,,\label{wr1}\\
\Xi_\sigma ({\bf p}) &=& \exp (-\Theta \,\hat {\bf p} \cdot {\bf J})
\Xi_\sigma ({\bf 0}) \,\label{wr2}
\end{eqnarray} \end{mathletters}
from the zero-momentum frame. $\Theta$ is the boost parameter,
$\tanh \,\Theta =\vert {\bf p} \vert/ E$, \,$\hat {\bf p} =
{\bf p}/ \vert {\bf p} \vert$, ${\bf p}$ is the 3-momentum of the particle,
${\bf J}$ is the angular momentum operator.
For a given representation the matrices ${\bf J}$ can be constructed. In
the Dirac case (the $(1/2,0)\oplus (0,1/2)$ representation) ${\bf J} =
{\bbox \sigma}/2$; in the $J=1$ case (the $(1,0)\oplus (0,1)$
representation) we can choose $(J_i)_{jk} = -i\epsilon_{ijk}$, etc. Hence,
we can explicitly calculate (\ref{wr1},\ref{wr2}), see Appendix for
$J=1/2, 1, 3/2, 2$ cases (cf. ref.~\cite{WHG}).

The task is now to obtain relativistic equations for higher spins.
Weinberg uses the following procedure (see also~\cite{Bar-Muz,Novozh}):

1) Firstly, he defined the scalar matrix
\begin{equation}
\Pi_{\sigma^\prime \sigma}^{(j)} (q) = (-)^{2j} t_{\sigma^\prime
\sigma}^{\quad \mu_1 \mu_2 \ldots \mu_{2j}} q_{\mu_1} q_{\mu_2}\ldots
q_{\mu_{2j}}
\end{equation}
for the $(J,0)$ representation of the Lorentz group ($q_\mu q_\mu =
-m^2$), with the tensor $t$ being defined by [1a,Eqs.(A4-A5)].
Hence,
\begin{equation}
D^{(j)} [\Lambda] \Pi^{(j)} (q) D^{(j)\,\dagger} [\Lambda] = \Pi^{(j)}
(\Lambda q)
\end{equation}
Since at rest we have $[{\bf J}^{(j)}, \Pi^{(j)} (m)] =0$, then according
to the Schur's lemma $\Pi_{\sigma\sigma^\prime}^{\quad (j)} (m) = m^{2j}
\delta_{\sigma \sigma^\prime}$. After the substitution of $D^{(j)}
[\Lambda]$ in Eq.  (5) one has
\begin{equation}
\Pi^{(j)} (q) = m^{2j} \exp (2\Theta \,\hat {\bf q}
\cdot {\bf J}^{(j)})\,.  \end{equation}
One can construct the analogous
matrix for the $(0,J)$ representation by the same procedure:
\begin{equation} \overline{\Pi}^{(j)} (q) = m^{2j} \exp (- 2\Theta
\hat{\bf q}\cdot {\bf J}^{(j)}) \,.  \end{equation}
Finally, by the direct
verification one has in the coordinate representation
\begin{mathletters}
\begin{eqnarray} \overline{\Pi}_{\sigma\sigma^\prime} (-i\partial)
\Phi_{\sigma^\prime} =m^{2j} \Xi_\sigma\,,\\ \Pi_{\sigma\sigma^\prime}
(-i\partial) \Xi_{\sigma^\prime} =m^{2j} \Phi_\sigma\,, \end{eqnarray}
\end{mathletters}
if $\Phi_\sigma ({\bf 0})$ and $\Xi_\sigma ({\bf 0})$ are
indistinguishable.\footnote{Later, this fact has been incorporated in the
Ryder book~\cite{Ryder}. Truely speaking, this is an additional postulate.
It is possible that the zero-momentum-frame $2(2J+1)$-component objects
(the 4-spinor in the $(1/2,0)\oplus (0,1/2)$ representation, the bivector
in the $(1,0)\oplus (0,1)$ representation, etc.) are connected by an
arbitrary phase factor~\cite{Dv-ff}.}

As a result one has
\begin{equation}
[ \gamma^{\mu_1 \mu_2 \ldots \mu_{2j}} \partial_{\mu_1} \partial_{\mu_2}
\ldots \partial_{\mu_{2j}} +m^{2j} ] \Psi (x) = 0\,,
\end{equation}
with the Barut-Muzinich-Williams covariantly-defined
matrices~\cite{Bar-Muz,Sankar}. For the spin-1 they are:
\begin{mathletters} \begin{eqnarray}
&&\gamma_{44} =\pmatrix{0&\openone\cr \openone&0\cr}\,,\quad
\gamma_{i4}=\gamma_{4i} = \pmatrix{0&iJ_i\cr
-iJ_i & 0\cr}\,,\\
&&\gamma^{ij} = \pmatrix{0&\delta_{ij} -J_i J_j - J_j J_i\cr
\delta_{ij} -J_i J_j - J_j J_i & 0\cr}\, .
\end{eqnarray}
\end{mathletters}
Later Sankaranarayanan and Good considered another version of this
theory~\cite{Sankar}. For the $J=1$ case they introduced the
Weaver-Hammer-Good sign operator, ref.~[5a], $m^{2} \rightarrow
m^{2}\, (i\partial/\partial t)/E$, which led to the different parity
properties of an antiparticle with respect to a {\it boson} particle.
Next,  Hammer {\it et al}~\cite{TH} introduced another higher-spin
equation. In the spin-1 case it is:
\begin{equation} [\gamma_{\mu\nu}
\partial_\mu \partial_\nu + \partial_\mu \partial_\mu -2m^2 ] \Psi^{(j=1)}
= 0\,  \end{equation}
(Euclidean metric is now used). In fact, they added the Klein-Gordon
equation to the Weinberg equation.

Weinberg considered massless cases too. He claimed that there is no
problem~[1b] to put $m\rightarrow 0$ in propagators and field functions
of the $(J,0)$ and $(0,J)$ fields. {\it But}, there are indeed problems for
the fields of the $(J/2,J/2)$-types, e.g., for the 4-vector potential.
The Weinberg theorem says~[1b,p.B885]: ``A massless particle operator
$a({\bf p},\lambda)$ of helicity $\lambda$ can {\it only} be used to
construct fields which transform according to representations $(A,B)$ such
that $B-A=\lambda$. For instance, a left-circularly polarized photon with
$\lambda =-1$ can be associated with $(1,0)$, $(3/2,1/2)$, $(2,1)$ fields
but {\it not} with the 4-vector potential $(1/2,1/2)$, at least
until we broaden our notion of what we mean by a Lorentz transformations".
He indicated that this is a consequence of the non-semi-simple structure
of the little group. In his book~\cite[\S 5.9]{Weinb}, he gave
additional details what did he mean in the above statement. Indeed, the
divergent terms of the 4-vector potential ($\lambda = \pm 1$) in the
$m\rightarrow 0$ limit may be removed by {\it gauge transformations}.

Another way of doing the massless limits (the Wigner-In\"on\"u
contraction $O(3)\rightarrow E(2)$, ref.~\cite{WI}, the infinite-momentum
limit) has been studied in~\cite{BK} and it will be presented by Prof.
Baskal in a separate talk.  They showed that one should take care to
consider gauge degrees of freedom here, and relations have been
found between spinors and gauge-(in)dependent parts of $A_\mu$ and
$F_{\mu\nu}$.

My present talk is, in fact, the extension of the results of the series
of the papers of the nineties, ref.~\cite{HPA}. One can add the
Klein-Gordon equation with arbitrary multiple factor to the Weinberg
equation. So, we study the generalized Weinberg-Tucker-Hammer equation
($J=1)$, which is written ($p_\mu = -i\partial/\partial x^\mu$):
\begin{equation}
[\gamma_{\alpha\beta}p_\alpha p_\beta +A p_\alpha p_\alpha +Bm^2 ]
\Psi =0\,.
\end{equation}
It has solutions with relativistic dispersion relations $E^2 -{\bf
p}^2 = m^2$, ($c=\hbar=1$) if
\begin{equation}
{B\over A+1} = 1\,, \qquad \mbox{or} \qquad{B\over A-1} =1\,.\label{se}
\end{equation}
This can be proven considering the algebraic equation
$Det [\gamma_{\alpha\beta} p_\alpha p_\beta +A p_\alpha p_\alpha +Bm^2 ]
=0$. It is  of the 12th order in $p_\mu$. Solving it with respect to
energy one obtains the conditions (\ref{se}).

Some particular cases are:

\begin{itemize}

\item
$A=0$, $B=1$. We obtain the Weinberg equation for $J=1$ with 3 solutions
of $E=+\sqrt{{\bf p}^{\,2} +m^2}$, 3 solutions of $E=- \sqrt{{\bf p}^{\,2}
+m^2}$, 3 solutions of $E=+\sqrt{{\bf p}^{\,2} -m^2}$ and
3 solutions of $E=-\sqrt{{\bf p}^{\,2} -m^2}$. The latter 6 solutions are
the tachyonic ones.

\item
$A=1$, $B=2$. We obtain the Tucker-Hammer equation\footnote{The tachyonic
version is
$[\gamma_{\alpha\beta}p_\alpha p_\beta + p_\alpha p_\alpha -2m^2 ]
\Psi^{tach} =0\,$; or its ``double"
is
$[\gamma_{\alpha\beta}p_\alpha p_\beta - p_\alpha p_\alpha +2m^2 ]
\Psi^{\prime\,tach} =0\,$.} for $J=1$, and we have only
solutions with $E=\pm \sqrt{{\bf p}^{\,2}+m^2}$.

\end{itemize}

One can try to find corresponding equations for antisymmetric tensor
(AST) fields of the 2nd rank. There are four interesting choices:

\begin{itemize}

\item
$\Psi_{(1)} = \pmatrix{{\bf E} +i{\bf B}\cr
{\bf E} -i{\bf B}\cr}$, where ${\bf E}_i=-iF_{4i}$, ${\bf B}_i = {1\over
2} \epsilon_{ijk} F_{jk}$\,.

\item
$\Psi_{(2)} = \pmatrix{{\bf B} -i{\bf E}\cr
{\bf B} +i{\bf E}\cr} = -i\gamma^5 \Psi_{(1)}$, where ${\bf
E}_i=-iF_{4i}$, ${\bf B}_i = {1\over 2} \epsilon_{ijk} F_{jk}$\,.

\item
$\Psi_{(3)} = \pmatrix{{\bf E} +i{\bf B}\cr
{\bf E} -i{\bf B}\cr}$, where ${\bf E}_i=-{1\over 2}\epsilon_{ijk}
\tilde F_{jk}$, ${\bf B}_i = -i\tilde F_{4i}$\,, {\it i.e.} the
corresponding vector and axial-vector fields are the components of the
{\it dual} tensor.

\item
$\Psi_{(4)} = \pmatrix{{\bf B} -i{\bf E}\cr
{\bf B} +i{\bf E}\cr}$, where ${\bf E}_i=-{1\over 2}\epsilon_{ijk}
\tilde F_{jk}$, ${\bf B}_i = -i\tilde F_{4i}$\,.

\end{itemize}

The first case has been considered in~[15c]. We obtained
\begin{equation}
\partial_\alpha\partial_\mu F_{\mu\beta}
-\partial_\beta \partial_\mu F_{\mu\alpha} +
{A-1\over 2} \partial_\mu \partial_\mu F_{\alpha\beta} - {B\over 2}
m^2 F_{\alpha\beta} = 0\,.\label{fc}
\end{equation}
In the Tucker-Hammer case $A=1$, $B=2$ we can recover the standard
equation for a massive AST field, which follows from the Proca
theory:\footnote{The procedure is simple. We express $A_\alpha = {1\over
m^2} \partial_\nu F_{\nu\alpha}$ and substitute it to the second Proca
equation $F_{\alpha\beta} = \partial_\alpha A_\beta - \partial_\beta
A_\alpha$.  Then, we multiply the resulting equation by $m^2$.}
\begin{equation}
\partial_\alpha \partial_\mu F_{\mu\beta} - \partial_\beta \partial_\mu
F_{\mu\alpha} = m^2 F_{\alpha\beta} \,.  \end{equation}
We also
noted~[15d]  that the massless limit of this theory does {\it not}
coincide with the Maxwell theory in full, while it contains the latter as
{\it a particular case}.  One should take precautions when considering the
massless limit of the Proca theory, see ref.~\cite{LG,Dv-rev2}. It is
possible to define various massless limits for the Proca-Duffin-Kemmer
theory for $J=1$. Another one is the Ogievetski\u{\i}-Polubarinov {\it
notoph}~\cite{OP} (which is also called as the Kalb-Ramond field in the US
literature~\cite{KR}). It was explicitly shown in~[15c] that the
Weinberg-Tucker-Hammer equations admit the non-transverse fields. For
instance,
\begin{mathletters}
\begin{eqnarray} \lefteqn{\left [E^2 -\vec
p^{\,2}\right ]({\bf E}+i{\bf B})^{\parallel} -m^2 ({\bf E}-i{\bf
B})^{\parallel} +\nonumber}\\ &+&\left [E^2 +\vec p^{\,2}- 2
E (\vec J\cdot\vec p)\right ] ({\bf E}+i{\bf B})^{\perp} - m^2 ({\bf
E}-i{\bf B})^{\perp} =0\,,\\ \lefteqn{\left [E^2 -\vec p^{\,2}\right
]({\bf E}-i{\bf B})^{\parallel} -m^2 ({\bf E}+i{\bf B})^{\parallel}
+\nonumber}\\ &+&\left [E^2 + \vec p^{\,2}+2 E (\vec J\cdot\vec
p)\right ]({\bf E}-i{\bf B})^{\perp} - m^2 ({\bf E}+i{\bf B})^{\perp}
=0 \,.
\end{eqnarray} \end{mathletters}
In the $m\rightarrow 0$ limit the statement remains to be true,
see also the review~\cite{Dv-rev2}.
For applications of the notoph concept to the boson mass generation
(including in the non-Abelian theories), see ref.~\cite{Am}.

The second case corresponds to the following equation for
the AST field:\footnote{See also hep-ph/9304243 concerning with a
unusual Weinberg-like theory. However, the studied equation coincides
with that of Sankaranarayanan and Good~[10a].}
\begin{equation} \partial_\alpha \partial_\mu
F_{\mu\beta} - \partial_\beta \partial_\mu F_{\mu\alpha} - {A+1 \over 2}
\partial_\mu \partial_\mu F_{\alpha\beta} +{B\over 2} m^2 F_{\alpha\beta}
= 0\,.  \label{sc}
\end{equation}
In the Tucker-Hammer case ($A=1$, $B=2$) it has no solutions
compatible with the massive Proca theory for $J=1$. After the substitution
$\partial_\mu \partial_\mu \rightarrow m^2$ we remain with the equation
$\partial_\alpha \partial_\mu F_{\mu\beta} - \partial_\beta \partial_\mu
F_{\mu\alpha} =0$, which rather corresponds to the free Maxwell-like case.
Of course, in the massless case $m=0$ we have well-known solutions.

If we consider the Weinberg cases ($A=0$ and $B=1$, or, alternatively,
$A=0$, $B=-1$) the sets of solutions of the corresponding Weinberg
equations are naturally separated into the causal $(\partial_\mu
\partial_\mu \rightarrow +m^2)$ and the tachyonic $(\partial_\mu
\partial_\mu \rightarrow - m^2)$ ones.  Let us consider the solutions of
the first kind (originated from the Weinberg-equation solutions
$\Psi_{(1)}$ with the apparent parity eigenvalue $-1$).  The substitution
$(\partial_\mu \partial_\mu \rightarrow +m^2)$ into the tensor equation
(\ref{fc}) corresponds to the solutions of the Tucker-Hammer equation
(which are causal in the sense $E^2 -{\bf p}^{\,2}=m^2$).  In the case of
the Weinberg ``double" we can obtain the corresponding tachyonic version.

Now let us consider the solutions of the second kind (originated from
the Weinberg-equation solutions $\Psi_{(2)}$ with the apparent parity
eigenvalue $+1$). The substitution
$(\partial_\mu \partial_\mu \rightarrow -m^2)$ into the tensor equation
will lead to a tachyonic version of the Tucker-Hammer equation.
The causal substitution $(\partial_\mu \partial_\mu
\rightarrow + m^2)$ into the tensor equation corresponding to the Weinberg
``double" will lead to $\partial_\alpha \partial_\mu F_{\mu\beta} -
\partial_\beta \partial_\mu F_{\mu\alpha} =m^2$, {\it i.e.} the solutions
of the causal Tucker-Hammer equation can be recovered.

Indeed, if one has
\begin{equation}
[\gamma_{\mu\nu} p_\mu p_\nu +m^2] \Psi = 0\,,
\end{equation}
then

\begin{tabular}{|l|l|}
$P=-1$, $\Psi_{(1)} = \pmatrix{{\bf E} +i {\bf B}\cr
{\bf E}-i{\bf B}}$, &
$P=+1$, $\Psi_{(2)} = \pmatrix{{\bf B} -i {\bf E}\cr
{\bf B}+i{\bf E}}$, \\
$\partial_\alpha \partial_\mu F_{\mu\beta}
-\partial_\beta \partial_\mu F_{\mu\alpha}
= {1\over 2} (m^2 +\partial_\mu^2) F_{\alpha\beta}$.
&$\partial_\alpha \partial_\mu F_{\mu\beta}
-\partial_\beta \partial_\mu F_{\mu\alpha}
= {1\over 2} (-m^2 +\partial_\mu^2) F_{\alpha\beta}$.\\
If $\partial_\mu^2 \rightarrow m^2$ we obtain
the mapping & If $\partial_\mu^2 \rightarrow m^2$ we obtain
zero in \\
of the Tucker-Hammer equations, which
&the right-hand side.\\
possess only causal solutions&\\
If $\partial_\mu^2 \rightarrow -m^2$ (tachyonic)
we obtain & If $\partial_\mu^2
\rightarrow -m^2$ we obtain \\
zero in the right-hand side. & the mapping of the tachyonic
versions\\
&of the Tucker-Hammer equations:\\
&$[\gamma_{\mu\nu} p_\mu p_\nu \pm p_\mu p_\mu \mp 2m^2]\Psi_t=0$, \\
& which has only
$E=\pm \sqrt{{\bf p}^{\,2}-m^2}$.
\end{tabular}

\vspace*{5mm}

Similar analysis can be produced for another Weinberg equation
$[\gamma_{\mu\nu} p_\mu p_\nu -m^2] \Psi = 0$, or even for equations with
different choices of the $A$ and $B$ parameters.

So, the conclusion is: in the case of the Weinberg equation the states of
different parities can be considered to be naturally separated into the
set of causal states and the set of tachyonic states. For any other choice
of parameters $A$ and $B$ in the generalized Weinberg-Tucker-Hammer
equation, this is not so. Nevertheless, in those cases the equations can
give {\it causal} states with {\it various} masses.\footnote{An
interesting case is $B=8$ (hence, $A=B-1=7$; an alternative case is
$B=-8$, $A=-7$).  In this case we can describe {\it various} mass states
which are connected by the relation $$m^{\prime^{\,2}} = {4\over 3} m^2.$$
We start from
\begin{equation} [\gamma_{\mu\nu} p_\mu p_\nu +(B-1) p_\mu
p_\mu + B m^2]\Psi =0 \end{equation}
with $B=8$. In this case
\begin{mathletters} \begin{eqnarray} \partial_\alpha \partial_\mu
F_{\mu\beta} - \partial_\beta \partial_\mu F_{\mu\alpha} &=& -3
(\partial_\mu \partial_\mu) F_{\alpha\beta} +4m^2 F_{\alpha\beta} \quad
(\mbox{originated from}\,\,\, P=-1)\,,\\
\partial_\alpha \partial_\mu
F_{\mu\beta} - \partial_\beta \partial_\mu F_{\mu\alpha} &=&4
(\partial_\mu \partial_\mu) F_{\alpha\beta} -4m^2 F_{\alpha\beta} \quad
(\mbox{originated from}\,\,\, P=+1)\,.  \end{eqnarray}
\end{mathletters}
If we consider $\partial_\mu^2 =m^2$ the first equation
will give us {\it only} causal solutions with mass $m$; the second one
will not give us any solutions which are compatible with the
massive Proca theory.
Let us consider also $\partial_\mu^2 = \kappa m^2$.
If we do not want to have Proca-like solutions from the first equation we
obtain
\begin{equation} (4-3\kappa) m^2 = 0 \quad \rightarrow
\kappa={4\over 3}\,.  \end{equation}
On using this $\kappa=4/3$ in the
second equation we observe that the right-hand side will become to be
equal
\begin{equation} (4\kappa - 4) m^2 = {4\over 3} m^2\,,
\end{equation}
{\it i.e.} a compatible solution of the mapping of the Tucker-Hammer
equation
\begin{equation} \partial_\alpha \partial_\mu F_{\mu\beta} -
\partial_\beta \partial_\mu F_{\mu\alpha} = {4\over 3} m^2 \end{equation}
for a causal massive particle with the squared mass ${4\over 3} m^2$.  For
other choices of parameters $A$ and $B$ we shall have other relation
between mass states.  However, the general conclusion is preserved. In
fact, we shall always have several {\it causal} solutions with mass
splitting from Eqs. (20).}$^,$\footnote{The second-order equation with
additional parameters has been considered by Barut~\cite{Bar} in an
analog of the $(1/2,0)\oplus (0,1/2)$ representation. See the discussion
in ref.~\cite{Dvijtp}}

It appears that similar equations for {\it dual} tensors follow from the
corresponding Weinberg-Tucker-Hammer formulation. For instance, the
third case ($\Psi^{(3)} = \mbox{column} ({\bf E}+i {\bf B} \quad
{\bf E} - i{\bf B})$, but ${\bf E}_i = -{1\over 2} \epsilon_{ijk} \tilde
F_{jk}$, ${\bf B}_i = -i \tilde F_{4i}$,
\begin{equation}
[\gamma_{\mu\nu} p_\mu p_\nu - p_\mu p_\mu - 2m^2]\Psi^{(3)} =0\,
\end{equation}
corresponds to
\begin{equation}
m^2 \tilde F_{\alpha\beta} = \partial_\alpha \partial_\mu
\tilde F_{\mu\beta} - \partial_\beta \partial_\mu \tilde F_{\mu\alpha}\,.
\end{equation}
The fourth case is ($\Psi^{(4)} = \mbox{column} ({\bf B}-i {\bf E}, \quad
{\bf B} + i{\bf E})$, ${\bf E}_i$ and ${\bf B}_i$
are the components of the dual tensor)
\begin{equation}
[\gamma_{\mu\nu} p_\mu p_\nu - p_\mu p_\mu - 2m^2]\Psi^{(4)} =0\,.
\end{equation}
It corresponds to
\begin{equation}
[\partial_\mu \partial_\mu - m^2] \tilde F_{\alpha\beta} = \partial_\alpha
\partial_\mu \tilde F_{\mu\beta} - \partial_\beta \partial_\mu \tilde
F_{\mu\alpha}\,.
\end{equation}

I would like to make the following comments:
\begin{itemize}

\item
In fact, we assumed that $\Psi^{(3)}$ and $\Psi^{(4)}$ satisfy
the $\gamma^5$-transformed equations.
The $\gamma^5$ transformation on the Tucker-Hammer equation thus may
induce {\it dual} transformations on the corresponding parts of the AST
field.

\item
During the dual transformations, we
have
$$m^2 \tilde F_{\alpha\beta} = \partial_\alpha \partial_\mu \tilde
F_{\mu\beta} - \partial_\beta \partial_\mu \tilde F_{\mu\alpha}
\quad\rightarrow \quad [\partial_\mu \partial_\mu - m^2]  F_{\alpha\beta}
= \partial_\alpha \partial_\mu  F_{\mu\beta} - \partial_\beta \partial_\mu
F_{\mu\alpha}\,,$$
and
$$
[\partial_\mu \partial_\mu - m^2] \tilde F_{\alpha\beta} = \partial_\alpha
\partial_\mu \tilde F_{\mu\beta} - \partial_\beta \partial_\mu \tilde
F_{\mu\alpha} \quad\rightarrow\quad
m^2  F_{\alpha\beta} = \partial_\alpha \partial_\mu
F_{\mu\beta} - \partial_\beta \partial_\mu  F_{\mu\alpha}\,.
$$

\item
So, the general solutions can be presented, e.g.,
$$\Psi = c_1 \Psi^{(1)} +c_2 \Psi^{(2)},$$
but for the AST field
$$G_{\mu\nu} = c_1^\prime F_{\mu\nu}^{(1)} + c_2^\prime \tilde
F_{\mu\nu}^{(3)}$$
should be considered. Alternatively,
$$\Psi^\prime = c_3 \Psi^{(3)} +c_4 \Psi^{(4)},$$ but
$$G_{\mu\nu}^\prime = c_3^\prime F_{\mu\nu}^{(2)} + c_4^\prime \tilde
F_{\mu\nu}^{(4)}\,.$$

\end{itemize}

Now we are going to study the connections of the obtained equations for
the AST field with other formalisms for higher spins. The first case
corresponds to the Proca theory, which can be obtained~\cite{Lurie} from
the well-known Bargmann-Wigner formalism~\cite{BW}. How can the second
case be obtained from the Bargmann-Wigner formalism? In their
paper~\cite{BW} they claimed explicitly that they wanted to construct
$2J+1$ states. In order to find the complete mapping with our
consideration we need $2(2J+1)$ states! In fact, they did not take into
account {\it parity} matters in~\cite{BW}. Nevertheless, see Wigner
lectures of 1962, ref.~\cite{Wiglec}, and cf. with~\cite{Gelf}.

The standard Bargmann-Wigner formalism in application to the $J=1,3/2$
systems can be found in~\cite{Lurie}. The $J=1$ field is described by a
symmetric $4\otimes 4$-spinor satisfying the Dirac equation in every index.
It is easy to come to the Proca theory (and, hence, to the first form of
the equation for the AST field following from the WTH equation for the
field function with $P=-1$):  \begin{mathletters} \begin{eqnarray} \left [
i\gamma_\mu \partial_\mu +m \right ]_{\alpha\beta} \Psi_{\beta\gamma} &=&
0\,,\label{bw1}\\ \left [ i\gamma_\mu \partial_\mu +m \right
]_{\gamma\beta} \Psi_{\alpha\beta} &=& 0\,, \label{bw2} \end{eqnarray}
\end{mathletters}
if one has
\begin{equation} \Psi_{\left \{ \alpha\beta
\right \} } = (\gamma_\mu R)_{\alpha\beta} A_\mu +
(\sigma_{\mu\nu} R)_{\alpha\beta} F_{\mu\nu}\,,
\end{equation} with
\begin{equation} R\sim CP = e^{\i\varphi}
\pmatrix{\Theta&0\cr 0&-\Theta\cr}\,\quad
\Theta=\pmatrix{0&-1\cr
1&0\cr}
\end{equation} in the spinorial
representation of $\gamma$-matrices.\footnote{It is considered that
$(\gamma_\mu R)$ and $(\sigma_{\mu\nu} R)$ are the only (?) symmetric
matrices  of the complete set. One obtains \begin{mathletters}
\begin{eqnarray}
&&\partial_\alpha F_{\alpha\mu} = {m\over 2} A_\mu\,,\\
&& 2m F_{\mu\nu} = \partial_\mu A_\nu - \partial_\nu A_\mu\,.
\end{eqnarray}
\end{mathletters}
(In order to obtain this set one should add the equations
(\ref{bw1},\ref{bw2}) and compare functional coefficients before the
corresponding commutators, see~\cite{Lurie}).  After the corresponding
re-normalization $A_\mu \rightarrow 2m A_\mu$, we obtain the standard
textbook set:
\begin{mathletters} \begin{eqnarray} &&\partial_\alpha
F_{\alpha\mu} = m^2 A_\mu\,,\\ && F_{\mu\nu} = \partial_\mu A_\nu -
\partial_\nu A_\mu\,.  \end{eqnarray} \end{mathletters} }
The connections
of corresponding vector and tensor functions with 2-spinors has been
considered in detail in~\cite{BK}.  The Weinberg theory is essentially the
$2(2J+1)$ theory. The BW theory is essentially the $2J+1$ theory, {\it
i.e.}  the latter  does {\it not} take into account {\it parity}
properties of the corresponding $J=1$ states. In order to get the complete
correspondence between the two theories one should generalize the
Bargmann-Wigner theory.

I modified~\cite{DV-Adv,DV-BW} the BW formalism putting an additional term
in the expansion of the BW 2-rank symmetric spinor:
\begin{equation}
\Psi_{\{\alpha\beta\}} \rightarrow
\Psi_{\{\alpha\beta\}}^{old} + b (\gamma_5 \sigma_{\mu\nu} R)_{\alpha\beta}
A_{\mu\nu}\,.
\end{equation}
In fact, I formed a superposition of
\begin{equation}
F_{\mu\nu}\rightarrow aF_{\mu\nu} +b \tilde A_{\mu\nu}\,.
\end{equation}
In such a way, I obtained the {\it dual} Proca-Duffin-Kemmer theory, which
may be reduced in $m\rightarrow 0$ to the well-known {\it dual}
electrodynamics~\cite{dual}. In such a way the equations
for the AST field of the third kind can be obtained. However, in order to
get the equations for the AST of the second kind (or, of the fourth
kind) from the procedure found in the Luri\`e book~\cite{Lurie}, one
should modify the formalism in a different way.

\smallskip

I propose:
\begin{itemize}

\item
To consider the {\it generalized parity-violating}  Dirac
equations~\cite{Dasfu,DV-GM} and [4e]\,\footnote{The generalized
parity-violating Dirac equation can be obtained in the following
way~[4e,31]:
\begin{equation} (E^2 -c^2 \vec{\bf p}^{\,2})
I^{(2)}\Psi= \left [E I^{(2)} - c {\bbox \sigma}\cdot {\bf p} \right ]
\left [E I^{(2)} + c {\bbox \sigma}\cdot {\bf p} \right ] \Psi = \mu_2^2
c^4 I^{(2)}\Psi\,.\label{G1} \end{equation}
If one denotes $\Psi=\eta$ one can
define $\chi = {1\over \mu_1 c} (i\hbar {\partial\over \partial x_0} -
i\hbar {\bbox \sigma}\cdot {\bbox\nabla}) \eta$.  The corresponding set of
2-component equations is
\begin{mathletters} \begin{eqnarray} &&(i\hbar
{\partial \over \partial x_0} -i\hbar {\bbox \sigma}\cdot {\bbox \nabla})
\eta =\mu_1 c\chi\,,\\ &&(i\hbar {\partial \over \partial x_0} +i\hbar
{\bbox \sigma}\cdot {\bbox \nabla}) \chi ={\mu_2^2 c\over \mu_1}\eta\,.
\end{eqnarray}
\end{mathletters}
In the 4-component form  we have
\begin{eqnarray}
&&\pmatrix{i\hbar (\partial/\partial x_0) &
i\hbar {\bbox \sigma}\cdot {\bbox \nabla}\cr
-i\hbar {\bbox \sigma}\cdot {\bbox \nabla}&
-i\hbar (\partial/\partial x_0)}
\pmatrix{\chi+\eta\cr\chi-\eta} = {c\over 2}
\pmatrix{(\mu_2^2/\mu_1
+\mu_1)&
(-\mu_2^2/\mu_1 +
\mu_1)\cr
(-\mu_2^2/\mu_1 +
\mu_1)& (\mu_2^2/\mu_1
+\mu_1)\cr}\pmatrix{\chi+\eta\cr\chi-\eta\cr}\,,\nonumber\\
&&
\end{eqnarray}
which results in
\begin{equation}
\left [i\hbar \gamma^\mu \partial_\mu - {\mu_2^2 c \over \mu_1}
{1-\gamma_5 \over 2} -\mu_1 c {1+\gamma_5 \over 2}\right ]
\Psi = 0\,.
\end{equation}
Choosing the natural unit system ($c=\hbar=1$) and
re-denoting ${\mu_2^2 \over \mu_1} +\mu_1 \rightarrow 2m_1$ and ${\mu_2^2
\over \mu_1} - \mu_1 \rightarrow -2m_2$ we obtain the equation
(39) presented above.}
\begin{equation} [i\gamma_\mu\partial_\mu +m_1 +m_2
\gamma_5] \Psi =0\,.  \end{equation}
This equation can describe massive
causal, massless (provided that $m_1=\pm m_2$) and tachyonic states
depending on the choice of mass parameters $m_1$ and $m_2$, ref.~[30j,k].

\item
To generalize it taking into account the Barut ideas of the mass-splitting
Dirac-like equation~\cite{Bar}
\begin{equation}
[i\gamma_\mu \partial_\mu + a {\partial_\mu \partial_\mu \over m} -
\kappa] \Psi =0\,.
\end{equation}
Fixing the parameter $a$ such that the equation will lead to the {\it
classical} anomalous magnetic moment of the corresponding field, we can
have two massive spin-1/2 states which correspond to the electron and
muon masses ($m_e \neq m_\mu$). The method of calculating the third
massive state has also been given by Barut~[21b].

\item
Instead of the analysis of a direct product of the two Dirac 4-spinors,
one can form the direct product of a 4-spinor and its
charge-conjugate.\footnote{In the case of consideration of the usual Dirac
equation there is no any difference with the standard formulation since
the charge-conjugate spinor satisfies the same Dirac equation.} In the
case of the use of the parity-violating generalized Dirac equation this
$4\otimes 4$ field function will satisfy {\it different} equations in
different indices (the sign before $\gamma_5$ term is changed for the
charge-conjugate spinor).

\end{itemize}

Thus, the parity-violating generalized Dirac equations for the modified
BW spinor $\Psi_{\beta\gamma}$ are
\begin{mathletters}
\begin{eqnarray}
\left [ i\gamma_\mu \partial_\mu + a -b \Box + \gamma_5 (c- d\Box )
\right ]_{\alpha\beta} \Psi_{\beta\gamma} &=&0\,,\\
\left [ i\gamma_\mu
\partial_\mu + a -b \Box - \gamma_5 (c- d\Box ) \right ]_{\alpha\beta}
\Psi_{\gamma\beta} &=&0\,,
\end{eqnarray} \end{mathletters}
with $a$, $b$, $c$ and $d$
being some unknown at this time dimensional coefficients; $\Box$ is the
d'Alembertian; the Euclidean metric is again used.  So, we shall have an
additional {\it anti}-symmetric part of the $4\otimes 4$ spinor:
\begin{mathletters} \begin{eqnarray} (i\gamma_\mu \partial_\mu + a +b
\partial_\mu \partial_\mu)_{\alpha\beta} \Psi_{\{\beta\gamma\}}
+\gamma^5_{\alpha\beta} (c+d  \partial_\mu \partial_\mu)
\Psi_{[\beta\gamma ]} &=&0\,,\\ (i\gamma_\mu \partial_\mu + a +b
\partial_\mu \partial_\mu)_{\gamma\beta} \Psi_{\{\alpha\beta\}}
-\gamma^5_{\gamma\beta} (c+d  \partial_\mu \partial_\mu)
\Psi_{[\alpha\beta ]} &=&0\,, \end{eqnarray} \end{mathletters} and
\begin{mathletters}
\begin{eqnarray}
(i\gamma_\mu \partial_\mu + a +b \partial_\mu \partial_\mu)_{\alpha\beta}
\Psi_{\left [ \beta\gamma\right ] } +\gamma^5_{\alpha\beta} (c+d
\partial_\mu \partial_\mu) \Psi_{\left \{\beta\gamma\right \} } &=&0\,,\\
(i\gamma_\mu \partial_\mu + a +b \partial_\mu \partial_\mu)_{\gamma\beta}
\Psi_{\left [ \alpha\beta\right ] } -\gamma^5_{\gamma\beta} (c+d
\partial_\mu \partial_\mu) \Psi_{\left \{ \alpha\beta\right \} } &=&0\,,
\end{eqnarray} \end{mathletters}
After performing the standard BW
procedure used to find the ``old" $J=1$ and $J=0$ sets of equations we
obtain the following Proca-like equations:
\begin{mathletters}
\begin{eqnarray} &&\partial_\nu A_\lambda - \partial_\lambda A_\nu - 2(a
+b \partial_\mu \partial_\mu ) F_{\nu \lambda} =0\,,\\ &&\partial_\nu
F_{\nu \lambda} = {1\over 2} (a +b \partial_\mu \partial_\mu) A_\lambda +
{1\over 2} (c+ d \partial_\mu \partial_\mu) \tilde A_\lambda\,,
\end{eqnarray} \end{mathletters}
with additional constraints:
\begin{mathletters}
\begin{eqnarray}
&&i\partial_\lambda A_\lambda + ( c+d\partial_\mu \partial_\mu) \tilde \phi
=0\,,\\
&&\epsilon_{\mu\lambda\kappa\tau} \partial_\mu F_{\lambda\kappa} =0\,,\\
&&( c+ d \partial_\mu \partial_\mu ) \phi =0\,.
\end{eqnarray} \end{mathletters}
And, the analogues of the $J=0$ Duffin-Kemmer equations are:
\begin{mathletters}
\begin{eqnarray}
&&(a+b \partial_\mu \partial_\mu) \phi = 0\,,\\
&&i\partial_\mu \tilde A_\mu  - (a+b\partial_\mu \partial_\mu) \tilde
\phi =0\,,\\
&&(a+b\, \partial_\mu \partial_\mu) \tilde A_\nu + (c+d\,\partial_\mu
\partial_\mu) A_\nu + i (\partial_\nu \tilde \phi) =0\,,
\end{eqnarray}
\end{mathletters}
with additional constraints:
\begin{mathletters}
\begin{eqnarray}
&&\partial_\mu \phi =0\,\\
&&\partial_\nu \tilde A_\lambda - \partial_\lambda \tilde
A_\nu +2 (c+d\partial_\mu \partial_\mu ) F_{\nu \lambda} = 0\,.
\end{eqnarray}
\end{mathletters}
Since $\tilde A_\mu$ is the 4-potential which is related to the spin-0
state~\cite{Weinb,LG,Dv-rev2}, we have that in the parity-violating
framework spin states are mixed.  For higher-spin equations similar
conclusions have been derived by Kruglov {\it et al} and Moshinsky {\it et
al}~\cite{Krug-Mosh}.

After elimination of the 4-vector potential we obtain the equation for the
AST field of the 2nd rank:
\begin{equation}
\left [ \partial_\mu \partial_\nu F_{\nu\lambda} - \partial_\lambda
\partial_\nu F_{\nu\mu}\right ]   + \left [ (c^2 - a^2) - 2(ab-cd)
\partial_\mu\partial_\mu  + (d^2 -b^2)
(\partial_\mu\partial_\mu)^2 \right ] F_{\mu\lambda} = 0\,,
\end{equation}
which should be compared with our previous equations which follow from the
Weinberg-like formulation (\ref{fc},\ref{sc}):
\begin{mathletters}
\begin{eqnarray}
c^2 - a^2 \rightarrow {-Bm^2 \over 2}\,,&\qquad& c^2 - a^2 \rightarrow
+{Bm^2 \over 2}\,,\\
-2(ab-cd) \rightarrow {A-1\over 2}\,,&\qquad&
+2(ab-cd) \rightarrow {A+1\over 2}\,,\\
b=\pm d\,.&\qquad&
\end{eqnarray}
\end{mathletters}
(The latter condition serves in order to exclude terms $\sim \Box^2$).
Of course, these sets of algebraic equations have solutions in terms $A$
and $B$. We found them and restored the equations (\ref{fc},\ref{sc}).

Thus the procedure which we made is the following one:
{\it The Modified Bargmann-Wigner formalism} $\rightarrow$
{\it The AST field equations} $\rightarrow$ {\it The Weinberg-Tucker-Hammer
approach}.

\bigskip

Finally, the conclusions are:

\begin{itemize}

\item
We found the mapping between the Weinberg formalism for $J=1$ and the
antisymmetric tensor fields of 4 kinds:
\begin{mathletters}\begin{eqnarray}
&&\partial_\alpha\partial_\mu F_{\mu\beta}^{(1)}
-\partial_\beta\partial_\mu F_{\mu\alpha}^{(1)}
+ {A-1\over 2} \partial_\mu \partial_\mu F_{\alpha\beta}^{(1)}
-{B\over 2} m^2 F_{\alpha\beta}^{(1)} = 0\,,\\
&&\partial_\alpha\partial_\mu F_{\mu\beta}^{(2)}
-\partial_\beta\partial_\mu F_{\mu\alpha}^{(2)}
- {A+1\over 2} \partial_\mu \partial_\mu F_{\alpha\beta}^{(2)}
+{B\over 2} m^2 F_{\alpha\beta}^{(2)} = 0\,,\\
&&\partial_\alpha\partial_\mu \tilde F_{\mu\beta}^{(3)}
-\partial_\beta\partial_\mu \tilde F_{\mu\alpha}^{(3)}
- {A+1\over 2} \partial_\mu \partial_\mu \tilde F_{\alpha\beta}^{(3)}
+{B\over 2} m^2 \tilde F_{\alpha\beta}^{(3)} = 0\,,\\
&&\partial_\alpha\partial_\mu \tilde F_{\mu\beta}^{(4)}
-\partial_\beta\partial_\mu \tilde F_{\mu\alpha}^{(4)}
+ {A-1\over 2} \partial_\mu \partial_\mu \tilde F_{\alpha\beta}^{(4)}
-{B\over 2} m^2 \tilde F_{\alpha\beta}^{(4)} = 0\,.
\end{eqnarray}\end{mathletters}

\item
Their massless limits contain additional solutions comparing with the
Maxwell equations. This is related to the possible existence of the
Ogievetski\u{\i}-Polubarinov-Kalb-Ramond notoph (which, in turn, should be
related to the Higgs field).

\item
In a particular case ($A=0$, $B=1$) massive solutions of different
parities are naturally divided into the classes of causal and tachyonic
solutions.

\item
If we want to take into account the Weinberg-equation $J=1$ solutions of
different parity properties, the Bargmann-Wigner formalism, the Proca one
and the Duffin-Kemmer formalism are to be generalized.

\item
In the $(1/2,0)\oplus (0,1/2)$ representation it is possible to introduce
the {\it parity-violating} framework.

\item
Adding the Klein-Gordon equation to the $(J,0)\oplus (0,J)$ equations may
change the theoretical content even on the free field level.\footnote{It
was known that such an addition indeed changes the physical consequences,
{\it but after} switching interactions on.}

\item
Higher-spin equations actually may describe various spin and mass states.

\end{itemize}

\section*{Appendix. Transformation Operators $\exp (\pm \Theta \,
\hat {\bf p}\cdot {\bf J})$}

\begin{tabular}{ll}
Spin 0 & 1\\
Spin 1/2 & $[ E+m \pm {\bbox\sigma}\cdot {\bf p} ]/[2m (E+m)]^{1/2} $\\
Spin 1 & $[ m(E+m) \pm (E+m) ({\bf J}\cdot {\bf p}) + ({\bf J}\cdot {\bf
p})^2 ]/[m(E+m)]$\\
Spin 3/2 &$ \frac{[-(E+m)(E-5m)\mp (2/3)(E-13m) ({\bf J}\cdot {\bf p})+
4({\bf J}\cdot {\bf p})^2 \pm (8/3) (E+m)^{-1} ({\bf J}\cdot {\bf p})^3
]}{[32m^3 (E+m)]^{1/2}}$\\
Spin 2 &$\frac{[m^2 (E+m)^2 \mp (1/3)(E-4m)(E+m)^2
({\bf J}\cdot {\bf p}) -(1/6) (E-7m) (E+m) ({\bf J}\cdot {\bf p})^2 \pm
(1/3) (E+m) ({\bf J}\cdot {\bf p})^3 + (1/6) ({\bf J}\cdot {\bf p})^4
]}{[m^2 (m+E)^2]}$\\ \end{tabular}

\acknowledgments
I greatly appreciate useful information from
Profs.  S. Baskal, S. Kruglov, Y. S. Kim and
Z.  Oziewicz.  Mr. P. Fendley and Mr. S. Warner
surely deserve certain kind of unusual acknowledgments.
This work has been supported in part by
the Mexican Sistema Nacional de Investigadores and
the Programa de Apoyo a la Carrera Docente.

\end{document}